\documentclass[12pt]{article}
\usepackage{times}
\usepackage{colordvi}
\usepackage{here}
\usepackage{amssymb}
\usepackage{amsmath}
\usepackage{epsfig}
\makeatletter\@addtoreset{equation}{section}\makeatother
\def\theequation{\thesection.\arabic{equation}}

\newcommand{\IGN}[1]{}

\begin{document}

\mbox{}
\\[-20mm]
\mbox{}
\hfill
{\small \sl Published in: \underline{Neural Computation {\bf 18}, 446-469 (2006)}}
\\\\
\noindent\large\bf
\mbox{Magnification Control in Self-Organizing Maps and Neural 
}
\\
Gas
\\\\
\\
\noindent
{\normalsize\bf Thomas Villmann}
\\
\small
\sl   
Clinic for {Psychotherapy,} University of Leipzig,
04107 Leipzig,
\small
\sl   
 Germany\\
\\
{\normalsize\bf  Jens Christian Claussen} \\
\small
\sl   
Institute {of Theoretical Physics and Astrophysics,
\\
\small
\sl   
Christian-Albrecht University Kiel, 24098 Kiel, Germany}\\
\small
\sl
\\\\
\normalsize
\small\bf
We consider different ways to control the
magnification in self-organizing maps (SOM) and neural gas (NG). Starting
from early approaches of magnification control in vector quantization, we
then concentrate on different approaches for SOM and NG.
We show that three structurally similar approaches can be 
applied to both algorithms:
localized learning, 
concave-convex learning, and winner relaxing learning.
Thereby, the approach of 
concave-convex learning in SOM is extended to a
more general description, whereas the 
concave-convex learning for NG is new.
In general, the control mechanisms generate only slightly different behavior
comparing both neural algorithms. However, we emphasize that the NG
results are valid for any data dimension, whereas in the SOM case the results
hold only for the one-dimensional case.
\normalsize \rm
%
\section{Introduction\label{sec_introduction}}
Vector quantization is an important task in data processing, pattern
recognition and control 
(Fritzke, 1993;
Haykin, 1994;
Linde, Buzo, \& Gray,
 1980;
Ripley, 1996). 
A large number of different types have been discussed,
(for an overview, refer to 
Haykin, 1994;
Kohonen, 1995;
Duda \& Hart, 1973). 
Neural maps are a popular type of \emph{neural vector quantizers} that
are commonly used in, for example, data visualization, feature extraction,
principle component analysis, image processing, classification tasks,
and acceleration of common vector quantization
(de Bodt, Cottrell, Letremy, \& Verleysen, 2004). 
Well
known approaches are the Self-Organizing Map (SOM) 
(Kohonen, 1995), 
the
neural gas (NG) 
(Martinetz, Berkovich, \& Schulten, 1993), 
elastic net (EN) 
(Durbin \& Willshaw, 1987)
and generative topographic mapping (GTM) (Bishop,
Svens\'en, \& Williams 1998).

In vector quantization, data vectors $\mathbf{v\in 
\mathbb{R}
}^{d}$ are represented by a few codebooks or weight vectors $\mathbf{w}_{i}$%
, where $i$ is an arbitrary index. Several criteria exist to evaluate the
quality of a vector quantizer. The most common
 \clearpage\noindent
 one is the squared
reconstruction error. However, other quality criteria are also known, for
instance topographic quality for neighborhood preserving mapping approaches 
(Bauer \& Pawelzik, 1992;
Bauer, Der, \& Villmann,  1999),
 optimization of mutual information 
(Linsker, 1989) and other criteria (for an overview, see 
Haykin, 1994).
Generally, a faithful representation of the data space by the codebooks is
desired. This property is closely related to the so-called magnification,
which describes the relation between data and weight vector density for a
given model. The knowledge of magnification of a map is essential for
correct interpretation of its output 
(Hammer \& Villmann, 2003). 
In addition,
explicit magnification control is a desirable property of learning
algorithms, if depending on the respective application, only sparsely
covered regions of the data space have to be emphasized or, conversely,
suppressed. The magnification can be explicitly expressed for several vector
quantization models. Usually, for these approaches the magnification can be
expressed by a power law between the codebook vector density $\rho $ and the
data density $P$. The respective exponent is called \emph{magnification
exponent }or \emph{magnification factor}. As explained in more
detail below, the magnification is also related to other properties of the
map, for example, reconstruction error as well as mutual information. Hence,
control of magnification is influencing these properties too.

In biologically motivated approaches, magnification can also be seen in
the context of information representation in brains, for instance, in the
senso-motoric cortex 
(Ritter, Martinetz, \& Schulten, 1992). Magnification and its control can be
related to biological phenomena like the perceptual magnet effect, which
refers to the fact that rarely occurring stimuli are differentiated with
high precision whereas frequent stimuli are distinguished only in a rough
manner 
(Kuhl, 1991; Kuhl, Williams, Lacerda, Stevens, \& Lindblom, 1992). 
It is a kind of attention-based
learning with inverted magnification, that is, 
rarely occurring input samples
are emphasized by an increased learning gain (Der
\& Herrmann, 1992;
Herrmann, Bauer, \& Der, 1994). 
This effect is also beneficial in technical systems. In
remote-sensing image analysis, for instance, seldomly found ground cover
classes should be detected, whereas usual (frequent) classes with broad
variance should be suppressed
(Mer{\'e}nyi \& Jain, 2004;
 Villmann, Mer{\'e}nyi \& Hammer, 2003).
Another technical environment for magnification control is robotics for
accurate description of dangerous navigation states 
(Villmann \& Heinze, 2000).

In this article we concentrate on a general framework for magnification
control in SOM and NG. In this context, we briefly review the most important
approaches. One approach for SOM is generalized, and afterward, it is
transferred to NG. For this purpose, we first give the basic notations,
followed in section \ref{section_basic_magnification_control}
by a more detailed description of magnification and early
approaches related to the topic of magnification control, including a unified
approach for controlling strategies.
The magnification control approaches of SOM are described
according to the unified framework in
section \ref{section_mgn_control_som},
 whereby one of them is significantly extended.
The same procedure is applied to NG in 
section \ref{section_mgn_control_ng}. 
Again, one of the control approaches presented in
this section is new. A short discussion concludes the article.

\section{Basic Concepts and Notations 
in SOM and NG\label{section_basic_notations}}
{In general,} neural maps project data vectors $\mathbf{v}$ from a (possibly
high-dimensional) data manifold $\mathcal{D}\subseteq $\textbf{$\mathbb{R}$}$%
^{d}$ onto a set $A$ of neurons $i$, which is formally written as $\Psi _{%
\mathcal{D}\rightarrow A}:\mathcal{D}\rightarrow A$. Each neuron $i$ is
associated with a pointer $\mathbf{w}_{i}\in $\textbf{$\mathbb{R}$}$^{d}$,
all of which establish the set 
$\mathbf{W}=\left\{ \mathbf{w}_{i}\right\}_{i\in A}$. 
The mapping description is a winner-take-all rule, that is, a
stimulus vector $\mathbf{v}\in \mathcal{D}$ is mapped onto that neuron $s\in
A$ with the pointer $\mathbf{w}_{s}$ being closest to the actual presented
stimulus vector $\mathbf{v}$, 
\begin{equation}
\Psi _{\mathcal{D}\rightarrow A}:\mathbf{v}\mapsto s\left( \mathbf{v}\right)
=\mathop{\rm argmin}_{i\in A}\left\Vert \mathbf{v}-\mathbf{w}_{i}\right\Vert .
\label{argmin}
\end{equation}%
The neuron $s$ is called {\sl winner neuron}. The set $\emph{R}_{i}=\left\{ 
\mathbf{v}\in \mathcal{D}|\Psi _{\mathcal{D}\rightarrow A}\left( \mathbf{v}%
\right) =i\right\} $ is called the {\sl (masked) receptive field} of the neuron $i$%
. The weight vectors are adapted during the learning process such that the
data distribution is represented.

For further investigations, we  describe SOM and NG as our focused neural
maps in more detail. During the adaptation process a sequence of data points 
$\mathbf{v}\in \mathcal{D}$ is presented to the map with respect to the data
distribution $P\left( \mathcal{D}\right) $. Then the most proximate neuron $%
s $ according to equation (\ref{argmin}) is determined, and the pointer 
$\mathbf{w}_{s}$, as well as all pointers $\mathbf{w}_{i}$ of neurons in the
neighborhood of $s$, are shifted towards $\mathbf{v}$, according to 
\begin{equation}
\bigtriangleup \mathbf{w}_{i}=\epsilon h\left( i,\mathbf{v},\mathbf{W}%
\right) \left( \mathbf{v}-\mathbf{w}_{i}\right) .  \label{allg_lernen}
\end{equation}%
The property of \textquotedblleft being in the neighborhood of $s$%
\textquotedblright\ is represented by a neighborhood function $h\left( i,%
\mathbf{v},\mathbf{W}\right) $. The neighborhood function is defined as 
\begin{equation}
h_{\lambda }\left( i,\mathbf{v},\mathbf{W}\right) =\exp \left( -\frac{%
k_{i}\left( \mathbf{v},\mathbf{W}\right) }{\lambda }\right)  \label{h_trn}
\end{equation}%
for the NG, where $k_{i}\left( \mathbf{v},\mathbf{W}\right) $ yields the
number of pointers $\mathbf{w}_{j}$ for which the relation $\left\Vert 
\mathbf{v}-\mathbf{w}_{j}\right\Vert <\left\Vert \mathbf{v}-\mathbf{w}%
_{i}\right\Vert $ is valid 
(Martinetz {et al}., 1993); 
especially, we have 
$h_{\lambda }\left( s,\mathbf{v},\mathbf{W}\right) =1.0$. In case of SOM the
set $A$ of neurons has a topological structure usually chosen as a hypercube
or hexagonal lattice. Each neuron $i$ has a fixed position $\mathbf{r}\left(
i\right) $. The neighborhood function has the form 
\begin{equation}
h_{\sigma }\left( i,\mathbf{v},\mathbf{W}\right) =\exp \left( -\frac{%
\left\Vert \mathbf{r}\left( i\right) -\mathbf{r}\left( s\left( \mathbf{v}%
\right) \right) \right\Vert _{A}}{2\sigma ^{2}}\right) .  \label{h_som}
\end{equation}
 \clearpage\noindent
In contrast to the NG, the neighborhood function of SOM is
evaluated in the output space $A$ according to its topological structure.
This difference causes the significantly different properties of both
algorithms.
For the SOM there does not exist any
energy function such that the adaptation rule follows the gradient descent 
(Erwin, Obermayer, \& Schulten, 1992). 
Moreover, the convergence proofs are only valid for the
one-dimensional setting 
(Cottrell, Fort \& Pages, 1998,
Ritter {et al}., 1992). 
The
introduction of an energy function leads to different dynamics as in the EN 
(Durbin \& Willshaw, 1987) 
or new winner determination rule 
(Heskes 1999). 
The advantage of the SOM is the ordered topological structure of
neurons in $A$. In contrast, in the original NG, such an order is \ not
given. One can extend the NG to the topology representing network (TRN) such
that topological relations between neurons are installed during learning,
although generally they do not achieve the simple structure as in SOM lattices 
(Martinetz \& Schulten, 1994). 
Finally, the important advantage of the NG is that
the adaptation dynamic of the weight vectors follows a potential minimizing
dynamics
(Martinetz {et al}.,  1993).

\section{Magnification and Magnification Control in Vector Quantization\label%
{section_basic_magnification_control}}

\subsection{\normalsize Magnification in Vector Quantization.}
Usually vector quantization aims to minimize the reconstruction error 
$RE=\sum_{i}\int_{\emph{R}_{i}}\left\Vert 
\mathbf{v}-\mathbf{w}_{i}\right\Vert ^{2}P
\left( \mathbf{v}\right) d\mathbf{v}$. 
However, other
quality criteria are also known, for instance, topographic quality
(Bauer \& Pawelzik, 1992;
Bauer {et al}.,  1999).
More generally, one can consider the
generalized distortion error,
\begin{equation}
E_{\gamma }=\int_{\mathcal{D}}\left\Vert \mathbf{w}_{s}-\mathbf{v}%
\right\Vert ^{\gamma }P\left( \mathbf{v}\right) d\mathbf{v.}
\label{generalized_distortion_error}
\end{equation}%
This error is closely related to other properties of the (neural) vector
quantizer. One important property is the achieved weight vector density $%
\rho \left( \mathbf{w}\right) $ after learning in relation to the data
density $P\left( \mathcal{D}\right) $. Generally, for vector quantizers one
finds the relation 
\begin{equation}
P\left( \mathbf{w}\right) \propto \rho \left( \mathbf{w}\right) ^{\alpha }
\label{magnification_law_general}
\end{equation}%
after the converged learning process (Zador 1982). The exponent $\alpha $
is called \emph{magnification exponent} or \emph{magnification factor}. The
magnification is coupled with the generalized distortion error (\ref%
{generalized_distortion_error}) by 
\begin{equation}
\alpha =\frac{d}{d+\gamma }  \label{equation_relation_magnification_to_error}
\end{equation}
 \clearpage\noindent
\begin{table}[H]
\caption{\small 
Magnification of Different Neural Maps  and Vector Quantization
Approaches. 
\label{tab_magnification_of_neural_maps}}
\vspace*{-2ex}
\begin{center}
\begin{tabular}{lcl}
\hline
Model & Magnification & Reference \\
\hline
Elastic 
net
& $1+\frac{\kappa }{\sigma ^{2}}\frac{1}{P\tilde{\rho}}$ & $\text{%
Claussen and Schuster (2002)}$ 
\\[-2mm] 
&  &  
\\[-2mm]
SOM & $\frac{1+12M_{2}\left( \sigma \right) }{3+18M_{2}\left( \sigma \right) 
}$ & $\text{Dersch and Tavan (1995)}$ 
\\[-2mm] 
&  &  
\\[-2mm]
Linsker network & $1$ & $\text{Linsker (1989)}$ 
\\[-2mm] 
&  &  
\\[-2mm]
LBG & $\frac{d}{d+2}$ & $\text{Zador (1982)}$ 
\\[-2mm] 
&  &  
\\[-2mm]
FSCL & $\frac{3\beta +1}{3\beta +3}$ & $\text{Galanopoulos and Ahalt (1996)}$ 
\\[-2mm] 
&  &  
\\[-2mm] 
NG & $\frac{d}{d+2}$ & $\text{Martinetz {et al}.\ (1993)}$ 
\\[-2mm] 
&  &  
\\[-2mm] 
\hline
\end{tabular}
\begin{minipage}{0.9\textwidth}
\small
Note: For SOM, $M_{2}\left( \sigma \right) $ denotes the 2nd
normalized moment of the neighborhood function depending on the neighborhood
range $\sigma $.
\end{minipage}
\end{center}
\normalsize
\end{table}
\noindent
where $d$ is the \emph{intrinsic} or 
{Hausdorff} dimension\footnote{%
Several approaches are known to estimate the Hausdorff dimension of data,
often called \emph{intrinsic dimension}. One of the best known methods is the
Grassberger-Procaccia-analysis (GP) 
(Grassberger \& Procaccia, 1983; Takens, 1985). 
For GP,
there is a large number of investigations of statistical properties
(e.g., Camastra and Vinciarelli, 2001;
Eckmann and Ruelle, 1992;
Liebert, 1991;
Theiler, 1990). 
For a
neural network approach of intrinsic dimension estimation (based on NG),
also in comparison to GP, we refer to 
Bruske and Sommer (1998),
Camastra \& Vinciarelli (2001),
Villmann, Hermann and Geyer (2000),
Villmann (2002), and 
Villmann {et al}.\ (2003).
}  
of the data. Beginning with the pioneering work of Amari (1980), which 
investigated a {resolution-density relation} of map formation in a neural
field model and extended the approach of 
Willshaw and 
{von der Malsburg (1976),
for several neural
map and vector quantizer approaches 
the} 
magnification relation has been
considered, including the investigation of the relation between data and
model density.

Generally, different magnification factors are obtained for different vector
quantization approaches. An overview of several important models with known
magnification factors is given in 
Table \ref{tab_magnification_of_neural_maps}. 


For the usual SOMs, mapping a one--dimensional input space onto a chain of
neurons,
\begin{equation}
\alpha _{SOM}=\frac{2}{3}
\end{equation}%
holds in the limit $1\ll \sigma \ll N$ 
(Ritter \& Schulten, 1986). 
For small values of
neighborhood range $\sigma $, the neighborhood ceases to be of influence, and
the magnification rate approaches the value $\alpha =\frac{1}{3}$ 
(Dersch \& Tavan, 1995). 
The influence of different types of neighborhood function was
studied in detail for SOMs in
Dersch and Tavan (1995), which extends the early
works 
of Luttrell (1991) and Ritter (1991). 
The magnification depends on the
second normalized moment $M_{2}$ of the neighborhood function, which itself
is determined by the neighborhood range $\sigma $. Van Hulle (2000) extensively
discussed the influence of kernel approaches in SOMs.
Results for magnification of discrete
SOMs can be found in 
Ritter (1989) 
and 
Kohonen (1999). These latter
problems and approaches will not be further addressed here.

According to equations (\ref{equation_relation_magnification_to_error}) and 
(\ref{generalized_distortion_error}), the SOM minimizes the somewhat exotic 
$E_{\frac{1}{2}}$ distortion error, whereas the NG minimizes the usual $E_{2}$-error.

Further, we can observe interesting relations to information-theoretic
properties of the mapping: The information transfer realized by the mapping $%
\Psi _{\mathcal{D}\rightarrow A}$, in general, is not independent of the
magnification of the map (Zador, 1982). It has been derived that for an
optimal information transfer realizing vector quantizer (or a neural map in
our context), the relation $\alpha =1$ holds 
(Brause, 1992). A vector
quantizer designed to achieve an optimal information transfer is the 
{Linsker network (Linsker, 1989; see 
Table \ref{tab_magnification_of_neural_maps}), 
or the optimal coding network approach
proposed by Brause (1994).
}
\vspace*{5mm}
\subsection{\normalsize Magnification Control in Vector Quantization: 
A General
Framework.\label{subsec_general_framework_of_control}}
\vspace*{-2mm}
As pointed out in
section \ref{sec_introduction},
 different application tasks may require
different magnification properties of the vector quantizer, 
that is, the
magnification should be controlled. 
Straightforwardly, magnification control
means changing the value of the magnification factor $\alpha $ for a given
vector quantizer by manipulation of the basic approach.

Consequently, the question is, How one can impact the magnification factor
to achieve an \emph{a priori} chosen magnification factor? We further
address this topic in the following.
First, we review results from the literature and put them into a general
framework.

The first approaches to influence the magnification of a vector quantizer
are models of 
{\emph{conscience learning},}  characterized by a modified
winner determination. The algorithm by DeSieno (1988) and the
frequency sensitive competitive learning (FSCL) 
(Ahalt, Krishnamurty, Chen, \& Melton, 1990)
 belong to
this algorithm class. Originally, these approaches were proposed for
equalizing the winner probability of the neural units in SOM. However, as
the neighborhood relation between neurons is not used in this approach, it
is applicable to each vector quantizer based on winner-take-all learning. To
achieve the announced goal, in the DeSieno model, a bias term $B$ is inserted
into the winner determination rule, equation (\ref{argmin}), such that%
\begin{equation}
\Psi _{\mathcal{D}\rightarrow A}:\mathbf{v}\mapsto s\left( \mathbf{v}\right)
=\mathop{\rm argmin}_{i\in A}\left( \left\Vert \mathbf{v}-\mathbf{w}%
_{i}\right\Vert -B\right)   \label{argmin_desieno}
\end{equation}%
with the bias term $B=\gamma \left( \frac{1}{N}-p_{i}\right) $, and $p_{i}$
is the actual winning probability of the neuron $i$. The algorithm converges
such that the winning probabilities of all neurons are equalized, which is
related to a maximization of the entropy,
and, hence, the resulted
magnification is equal to the unity. However, an arbitrary magnification can
not be achieved. Moreover, as pointed out in 
van Hulle (2000), the
algorithm shows unstable behavior. FSCL modifies the selection criterion
for the best-matching unit by a fairness term $F$, which is a function of
the winning frequency $\omega _{i}$ of the neurons. Again, the winner
determination is modified:
\begin{equation}
\Psi _{\mathcal{D}\rightarrow A}:\mathbf{v}\mapsto s\left( \mathbf{v}\right)
=\mathop{\rm argmin}_{i\in A}\left( F\left( \omega _{i}\right) \left\Vert 
\mathbf{v}-\mathbf{w}_{i}\right\Vert \right).
\end{equation}
As mentioned above, originally it was defined to achieve an equiprobable
quantization too. However, it was shown, this goal can not be achieved by
the original version (Galanopoulos \& Ahalt, 1996; van Hulle, 2000). 
Yet for one-dimensional data, any given $\gamma $-norm error criterion, 
equation (\ref%
{generalized_distortion_error}),
can be minimized by a specific choice of the
fairness function: if $F\left( \omega _{i}\right) $ is taken as 
\begin{equation}
F\left( \omega _{i}\right) =\left( \omega _{i}\right) ^{\xi }
\end{equation}%
for the one-dimensional case a magnification $\alpha _{FSCL}=\frac{3\beta +1%
}{3\beta +3}$ is achieved, being equivalent to $\gamma =\frac{2}{3\beta +1}$ 
(Galanopoulos \& Ahalt, 1996). The difficulties of transferring the 
one-dimensional result to higher dimensions are, however, as prohibitive as in
SOM.

We now study control possibilities to achieve \emph{arbitrary magnification},
focusing on SOM and NG by \emph{modification of the learning rule}. We
emphasize again that for SOM, the results hold only for the one-dimensional
case, whereas for NG, the more general case of arbitrary dimensionality is
valid. Thus, the following direction of modifications of the general
learning rule, equation (\ref{allg_lernen}),
\begin{equation*}
\bigtriangleup \mathbf{w}_{i}=\epsilon h\left( i,\mathbf{v},\mathbf{W}%
\right) \left( \mathbf{v}-\mathbf{w}_{i}\right),
\end{equation*}
can serve as a general framework:
\begin{enumerate}
\item \emph{Localized learning}: Introduction of a multiplicative factor by
a local learning rate $\epsilon _{i}$
\item \emph{Winner-relaxing learning}: Introduction of winner relaxing by
adding a winner-enhancing (relaxing) term $R$
\item 
\emph{{Concave-convex} learning}: Scaling of the learning shift by
powers $\xi $ in the factor $\left( \mathbf{v}-\mathbf{w}_{i}\right) ^{\xi }$
\end{enumerate}
These three directions serve as axes for a taxonomy in the following
section. We focus on SOM and NG as popular neural vector quantizers. We
explain, expand and develop the respective methodologies of magnification
control for these models. The localized and the winner relaxing
learning for SOM and NG are briefly reviewed. In particular, localized
learning for SOM was published in 
Bauer, Der, and  Herrmann, 
(1996)
 whereas winner relaxing
learning for both SOM and NG and localized learning in NG were 
previously developed by
the authors
(Claussen, 2003, 2005;
Claussen \& Villmann, 2003a;
Villmann, 2000). 
The 
{concave-convex
} learning for SOM is extended here to a
more general approach compared to its origins 
(Zheng \& Greenleaf, 1996). The
{concave-convex
} 
learning for NG is new too.

\section{Controlling the Magnification in SOM\label{section_mgn_control_som}}
Within the general framework outlined in section 
\ref{subsec_general_framework_of_control}, we now consider the three learning
rule modifications for SOM.

\subsection{\normalsize Insertion of a Multiplicative Factor: Localized Learning.}
The first choice is to add a factor in the SOM learning rule. 
An
established
realization is the \emph{localized learning}, the biological motivation of
which is the perceptual magnet effect 
(Bauer {et al}., 1996). 
For this purpose, an
adaptive local learning step size $\epsilon _{s\left( \mathbf{v}\right) }$
is introduced in equation (\ref{allg_lernen}) such that the new adaptation rule reads
as 
\begin{equation}
\bigtriangleup \mathbf{w}_{i}=\epsilon _{s\left( \mathbf{v}\right)
}h_{\sigma }\left( i,\mathbf{v},\mathbf{W}\right) \left( \mathbf{v}-\mathbf{w%
}_{i}\right)  \label{som_bdh_lernen}
\end{equation}%
where $s\left( \mathbf{v}\right) $ is being the best-matching neuron with
respect to equation (\ref{argmin}). The local learning rates $\epsilon _{i}=\epsilon
\left( \mathbf{w}_{i}\right) $ depend on the stimulus density $P$ at the
position of their weight vectors $\mathbf{w}_{i}$ via 
\begin{equation}
\left\langle \epsilon _{i}\right\rangle =\epsilon _{0}P\left( \mathbf{w}%
_{i}\right)^{m},  \label{som_bdh_eps_ansatz}
\end{equation}%
where the brackets $\left\langle \ldots \right\rangle $ denote the average
in time. This approach leads to the new magnification law, 
\begin{equation}
\alpha _{localSOM}^{\prime }=\alpha _{SOM}\cdot \left( m+1\right),
\label{som_mgn_alpha_bdh}
\end{equation}%
where $\ m$ appears to be an explicit control parameter 
(Bauer
et al.,
 1996).
Hence, an arbitrary predefined magnification can be achieved.

In applications, one has to estimate the generally unknown data
distribution $P$, which may lead to numerical instabilities of the control
mechanism (van Hulle, 2000).

\subsection{\normalsize Winner-Relaxing SOM and Magnification Control.}
Recently, a new approach for magnification control of the SOM by a
generalization 
(Claussen, 2003, 2005) 
of the 
{winner-relaxing} modification 
(Kohonen, 1991) 
was derived, giving a control
scheme, which is independent of the shape of the data distribution 
(Claussen 2005). 
We refer to this algorithm as WRSOM.

In the original winner-relaxing SOM, an additional term occurs in learning
for the winning neuron only, implementing a relaxing behavior. The relaxing
force is a weighted sum of the difference between the weight vectors and the
input according to their neighborhood relation. The relaxing term was
introduced to obtain a learning dynamic for SOM according to an average
reconstruction error taking into account the effect of shifting Voronoi
borders.

{The original learning rule is added by a winner relaxing term $R\left( \mu
,\kappa \right) $ as}
\begin{equation}
\bigtriangleup \mathbf{w}_{i}=\epsilon h_{\sigma }\left( i,\mathbf{v},%
\mathbf{W}\right) \left( \mathbf{v}-\mathbf{w}_{i}\right) +R\left( \mu
,\kappa \right),   \label{som_winner_relaxing_learning_allg}
\end{equation}%
with $R\left( \mu ,\kappa \right) $ being%
\begin{eqnarray}
R\left( \mu ,\kappa \right)  &=&\left( \mu +\kappa \right) \left( \mathbf{v}-%
\mathbf{w}_{i}\right) \delta _{is}  \label{som_winner_relaxing_term} \\
&&-\kappa \delta _{is}\sum_{j}h_{\sigma }\left( j,\mathbf{v},\mathbf{W}%
\right) \left( \mathbf{v}-\mathbf{w}_{j}\right),   \notag
\end{eqnarray}%
depending on weighting parameters $\mu $ and $\kappa $. For $\mu =0$ and $%
\kappa =\frac{1}{2}$, the original {winner relaxing SOM} is obtained 
(Kohonen, 1991). 
Surprisingly, it has been shown that the magnification is
independent of $\mu $ 
(Claussen, 2003, 2005). 
Only the choice of $\kappa $
contributes to the magnification: 
\begin{equation}
\alpha _{WRSOM}^{\prime }=\frac{2}{\kappa +3}.
\label{som_mgn_alpha_winner_relaxing}
\end{equation}%
The stability range is $\left\vert \kappa \right\vert \leq 1$, which 
restricts the accessible magnification range to $\frac{1}{2}\leq \alpha
_{WRSOM}^{\prime }\leq 1$. More detailed numerical simulations and stability
analysis can be found in Claussen (2005).

The advantage of winner relaxing learning is that no estimate of the
generally unknown data distribution has to be made, as required in the
local learning approach above.

\subsection{\normalsize {Concave-Convex} Learning.}
The third structural possibility for control according to our framework is
to apply \emph{{concave or convex} learning} in the learning rule. This
approach was  introduced in 
Zheng and Greenleaf (1996).
Here, we extend this
approach to a more general variant.

Originally, an exponent $\xi $ is introduced in the general learning rule
such that equation (\ref{allg_lernen}) now reads as 
\begin{equation}
\bigtriangleup \mathbf{w}_{i}=\epsilon h_{\sigma }\left( i,\mathbf{v},%
\mathbf{W}\right) \left( \mathbf{v}-\mathbf{w}_{i}\right) ^{\xi }
\label{som_zheng_learning}
\end{equation}%
 \clearpage\noindent
with 
\begin{equation}
\left( \mathbf{v}-\mathbf{w}_{i}\right) ^{\xi }\overset{def}{=}\left( 
\mathbf{v}-\mathbf{w}_{i}\right) \cdot \left\Vert \mathbf{v}-\mathbf{w}%
_{i}\right\Vert ^{\xi -1}.  \label{def_vector_potenz}
\end{equation}%
Thereby, two different possibilities are proposed: $\xi =\frac{1}{%
\kappa }$ with $\kappa >1$, $\kappa \in $\textbf{$\mathbb{N}$} and $\kappa $
is odd (\emph{convex learning}), or one simply takes $\xi >1$, $\xi \in $%
\textbf{$\mathbb{N}$} and $\xi $ is odd (\emph{concave learning}). This
gives the magnification 
\begin{eqnarray}
\alpha _{{concave/convex}SOM}^{\prime } &=&\frac{2}{\xi +2}
\label{som_mgn_alpha_zheng} \\
&=&\alpha _{SOM}\cdot \frac{3}{\xi +2}
\end{eqnarray}%
which allows an explicit magnification control. Yet this approach allows
only a rather rough control around $\xi =1$: the neighboring allowed values
are $\xi =\frac{1}{3}$ and $\xi =3$ corresponding to magnifications $\alpha
_{{concave/convex}SOM}^{\prime }=\frac{6}{7}$ and $\alpha
_{{concave/convex}SOM}^{\prime }=\frac{2}{5}$, respectively. Therefore,
greater flexibility would be of interest.

For this purpose, we are seeking for a generalization 
of both concave and
convex learning. As a more general choice we take $\xi $ to be real, 
that is, 
$\xi\in\mathbb{R}$. 
{If we do so, the same magnification 
equation (\ref{som_mgn_alpha_zheng}) is
obtained.
}
The proof of the magnification law is given in  
appendix A.
Obviously, the choices $\xi =\frac{1}{\kappa }$ and $\xi =\kappa >1$, $%
\kappa \in $\textbf{$\mathbb{N}$ }and $\kappa $ being odd as made in 
Zheng and Greenleaf (1996)
are special cases of the now general approach.

\begin{figure}[htpb]
\centerline{
\epsfig{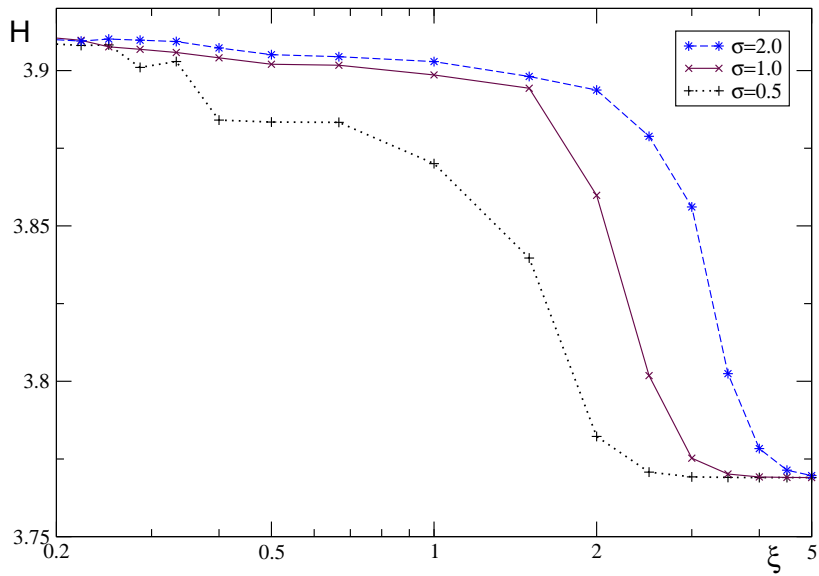}
}
\caption{\small Output entropy for concave and convex learning. 
An input density of $P(x)=\sin (\protect\pi x)$ was
presented to an one-dimensional chain of $N=50$ neurons after $10^{6}$
learning steps of stochastic sequential updating, averaged over $10^{5}$
inputs, and learning rate $\protect\epsilon =0.01$, fixed. 
\label{fig_som_concave_convex_entropy}}
\end{figure}

We
considered the numerical behaviour of the magnification control of the
{WRSOM
} 
using a one-dimensinal chain of $\ 50$ neurons. The data
distribution was chosen in agreement with 
Bauer et al. (1996)
as $P(x)=\sin (\pi
x).$The theoretical entropy maximum of the winning probabilities of the
neurons $p_{i}$ is $\sum_{i=1}^{N}p_{i}\log (p_{i})=\log (N)$ giving the
value $3.912$ for $N=50$. The results in dependence on $\xi $ for different
{neighborhood ranges $\sigma$ 
are 
depicted in Figure \ref{fig_som_concave_convex_entropy}.
}

According to the theoretical prediction, the output entropy is maximized for
small $\xi $, and for large $\xi$, an magnification exponent zero is reached
corresponding to an equidistant codebook without adaptation to the input
distribution. For $\sigma <1$, the turnover is shifted toward smaller
values of $\xi $, and for $\xi \ll 1,\sigma \ll 1$, fluctuations increase.

Further, as in the 
{WRSOM,
} the advantage of 
{concave-convex
} 
learning is that
no estimate of the generally unknown data distribution has to be made 
{as
before in localized learning.
}

\clearpage

\section{Magnification Control in Neural Gas \label{section_mgn_control_ng}}

In this section we transfer the ideas of magnification control in SOM to the
NG, keeping in mind the advantage that the results then are valid for any
dimension.

\subsection{\normalsize Multiplicative Factor - Localized Learning.}
The idea of \emph{localized learning} is now applied to NG 
(Herrmann \& Villmann 1997). 
Hence, we have the localized learning rule 
\begin{equation}
\bigtriangleup \mathbf{w}_{i}=\epsilon _{s\left( \mathbf{v}\right)
}h_{\lambda }\left( i,\mathbf{v},\mathbf{W}\right) \left( \mathbf{v}-\mathbf{%
w}_{i}\right),
\end{equation}
with $s\left(\mathbf{v}\right)$ 
again being the best-matching neuron
with respect to equation (\ref{argmin}) and $\epsilon_{s\left( \mathbf{v}\right) }$
is the local learning chosen as in 
equation (\ref{som_bdh_eps_ansatz}). This approach
gives a similar result as for SOM, 
\begin{equation}
\alpha _{localNG}^{\prime }=\alpha _{NG}\cdot \left( m+1\right),
\end{equation}
 \clearpage\noindent
\begin{figure}[hbtp]
\centerline{
\epsfig{file=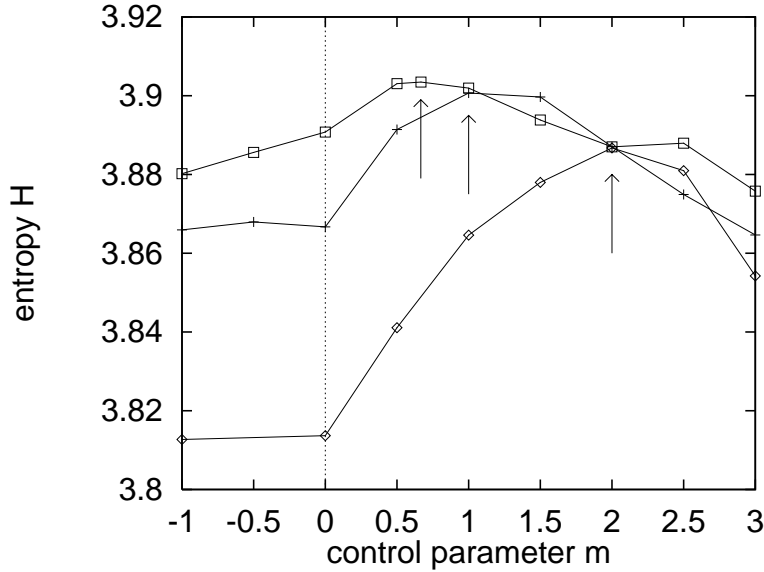,width=4.25in}
}
\caption{\small
{\sl Local learning} for NG:
Plot of the entropy $H$ for maps trained with different magnification
control parameters $m$ ($d=1$ ($\diamond $), $d=2$ ($+$), $d=3$ ($\Box $)).
The arrows indicate the theoretical values of $m$ ($m=2$, $m=1$, $m=2/3$,
resp.) which maximizes the entropy of the map.
\label{fig_ng_mgn_local_learning_result}}
\end{figure}

\noindent
and, hence, allows a magnification control 
(Villmann, 2000). However, we
have similar restrictions as for SOM: 
in actual applications one has to
estimate the generally unknown data distribution $P$.

The numerical study shows that
the approach can also be used to increase the mutual information of a map
generated by a NG 
(Villmann, 2000). As for WRSOM, we use a standard
setup as in 
Villmann (2000) of $50$ Neurons and $10^{7}$ training steps
with a probability density $P(x_{1}\ldots x_{d})=\prod_{i}\sin (\pi x_{i})$, 
$x\in \left[ 0,1\right] $, and with parameters $\lambda =1.5$ fixed and $%
\epsilon $ decaying from $0.5$ to $0.05$. The entropy of the resulting map
computed for an input dimension of $1$, $2$ and $3$ is plotted in 
Figure \ref{fig_ng_mgn_local_learning_result}.

\subsection{\normalsize Winner-Relaxing NG.}
The winner-relaxing NG (WRNG) was first studied in 
Claussen and Villmann (2003a).
According to the WRSOM approach, one uses an additive winner relaxing term $%
R\left( \mu ,\kappa \right) $ to the original learning rule: 
\begin{equation}
\bigtriangleup \mathbf{w}_{i}=\epsilon h_{\lambda }\left( i,\mathbf{v},%
\mathbf{W}\right) \left( \mathbf{v}-\mathbf{w}_{i}\right) +R\left( \mu
,\kappa \right),   \label{ng_winner_relaxing_learning}
\end{equation}
 \clearpage\noindent
with $R\left( \mu ,\kappa \right) $ being as in 
equation (\ref{som_winner_relaxing_term}). 
The resulting WRNG-magnification for
small neighborhood values $\lambda $ with $\lambda \rightarrow {}0$ but not
vanishing is given by 
Claussen and Villmann (2005): 
\begin{equation}
{\alpha }_{WRNG}^{\prime }=\frac{1}{1-\kappa }\frac{d}{d+2}.
\end{equation}%
Thereby, the magnification exponent appears to be independent of an
additional diagonal term (controlled by $\mu $) for the winner the same
as in WRSOM; again $\mu =0$ is the usual setting. If the same stability
borders $|\kappa |=1$ of the WRSOM also apply here, one can expect to
increase the NG exponent by positive values of $\kappa $, or to lower the NG
exponent by factor $1/2$ for $\kappa =-1$.

However, one has to be cautious when transferring the $\lambda \rightarrow
{}0$ result obtained above (which would require to increase the number of
neurons as well) to a realistic situation where a decrease of $\lambda $
with time will be limited to a final finite value to avoid the stability
problems found in 
Herrmann and Villmann (1997). 
For a finite $\lambda $ the maximal
coefficient $h_{\lambda }$ that contributes to the averaged learning shift
is given by the prefactor of the second but one winner, which is given by $%
\mathrm{e}^{\lambda }$ 
(Claussen \& Villmann, 2005).
For the NG, however, the
neighborhood is def\hspace*{0.03em}{}ined by the rank list. As the winner
term of the NG is not present in the winner relaxing term (for $\mu =0$), all
terms share the factor $\mathrm{e}^{-\lambda }$ by $h_{\lambda }(k)=\mathrm{e%
}^{-\lambda }h_{\lambda }(k-1)$ which indicates that in the discretized
algorithm $\kappa $ has to be rescaled by $\mathrm{e}^{+\lambda }$ to agree
with the continuum theory. The numerical investigation indicates that this
prefactor applies for finite $\lambda $ and number of neurons. The scaling
of the position of the entropy maximum with input dimension is in agreement
with theory, as well as the prediction of the opposite sign of $\kappa $
that has to be taken to increase mutual information.

Numerical studies show that  winner-relaxing learning
can also be used to increase the mutual information 
{of a NG vector
quantization.
} 
The entropy shows a dimension-dependent maximum approximately
at $\kappa =\frac{2}{d+2}\mathrm{e}^{\lambda }$ (see Figure \ref%
{fig_ng_mgn_winner_relaxing_result}). In any case, within a broad range
around the optimal $\kappa $, the entropy is close to the maximum.

The advantage of the method is to be independent on estimation of the
unknown data distribution as the SOM equivalent WRSOM. Further, again as in
the WRSOM, the magnification of WRNG is independent in the first order on the
diagonal term, controlled by $\mu $. Numerical simulations have shown that
the contribution in higher orders is marginal 
(Claussen \& Villmann, 2003b). More
pronounced is the inf\hspace*{0.03em}{}luence of the diagonal term on
stability. According to the larger prefactor, no stable behavior has been
found for $|\mu |\geq {}1$, therefore $\mu =0$ is the recommended setting 
(Claussen \&Villmann, 2005).
\clearpage
\begin{figure}[htpb]
\centerline{
\epsfig{file=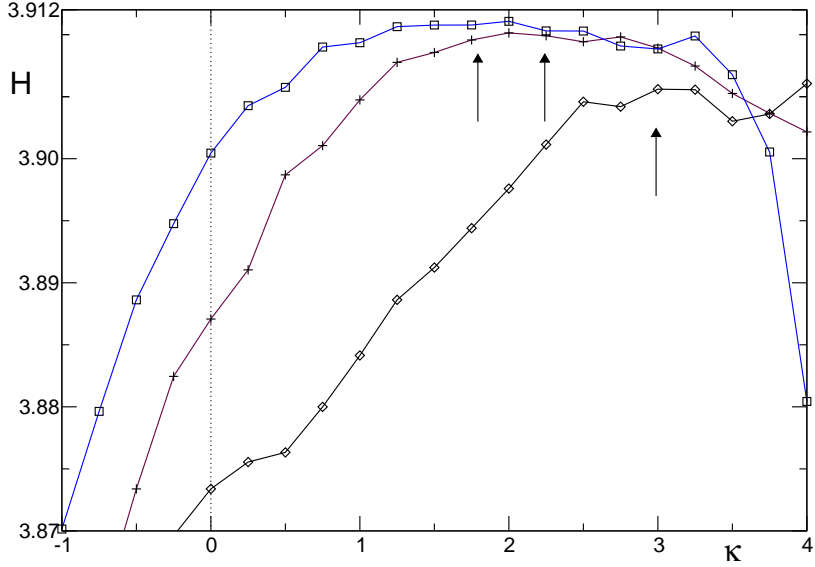,width=4.25in} 
}
\caption{\small {\sl Winner
relaxing learning} for NG: Plot of the entropy $H$ curves for varying values
of $\kappa $ for one-, two and three-dimensional data. The entropy
has the maximum if the magnification equals the unit 
(Zador 1982). 
The arrows indicate the $\kappa $-values for the respective data
dimensions.
\label{fig_ng_mgn_winner_relaxing_result}}
\end{figure}
\subsection{\normalsize {Concave-Convex} Learning.}
We now consider the third modification known from SOM, the 
{concave-convex
} 
learning approach but in its new, developed general variant, 
\begin{equation}
\bigtriangleup \mathbf{w}_{i}=\epsilon h_{\lambda }\left( i,\mathbf{v},%
\mathbf{W}\right) \left( \mathbf{v}-\mathbf{w}_{i}\right) ^{\xi},
\end{equation}%
with $\xi \in 
\mathbb{R}
$ and the definition (\ref{def_vector_potenz}). It is proved in the 
appendix B 
that the resulting magnification is 
\begin{equation}
\alpha_{{concave/convex}NG}^{\prime }=\frac{d}{\xi +1+d},
\end{equation}%
depending on the intrinsic data dimensionality $d$. This dependency is in
agreement with the usual magnification law of NG, which is also related to
the data dimension.

\begin{figure}[htpb]
\vspace*{-.1in}
\centerline{
\epsfig{file=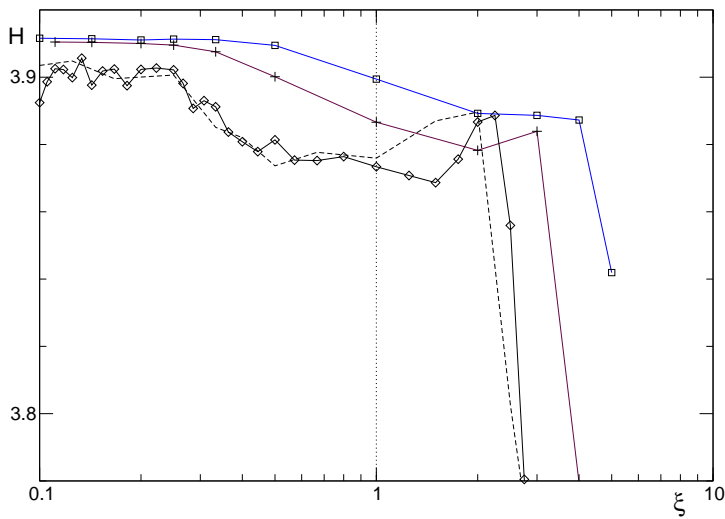,width=5.0in} 
}
\vspace*{-.15in}
\caption{\small {\sl {Concave-convex} learning} for NG: Plot of the entropy $H$ curves
for varying values of $\protect\xi $ for one-, two and three-dimensional
data. The entropy can be enhanced by convex learning in each case
 (dashed line: $%
d=1~$, with $10^{8}$ learning steps).
\label{fig_ng_mgn_concave_convex_learning_result}}
\end{figure}

The respective
numerical simulations with the parameter choice as before are given in 
Figure
\ref{fig_ng_mgn_concave_convex_learning_result}. In contrast to
{concave-convex
} 
SOM where $\alpha ^{\prime }=1$ can be achieved for large $%
\xi $, here $\alpha ^{\prime }$ is bounded by $\frac{d}{d+1}$;
information optimal learning is not possible in cases of low-dimensional
data.

\clearpage
\section{Discussion}
According to the given general framework, we studied three structurally
different approaches for magnification control in SOM and NG. All methods
are capable to control the magnification with more or less accuracy. Yet,
they differ in properties (e.g., stability range, density estimation). No
approach yet shows a clear advantage. The choice of the optimal algorithm
may depend on the particular problem and implementation constraints. In
particular, several problems occur in actual applications. First, in the
SOM case, all result are only valid for the one-dimensional case, because
all investigations are based on the usual convergence dynamic. However, the
SOM dynamics is analytically treatable only in the one-dimensional setting
and higher-dimensional cases that factorize. Moving away from these special
cases causes a systematic shift in magnification control, as 
numerically shown in 
Jain and Mer{\'e}nyi (2004).
In actual
applications, a quantitative comparison with theory is quite limited due to
several influences which are not easily tractable. First, the data density
has to be estimated, which is generally difficult 
(Mer{\'e}nyi \& Jain, 2004);
second, the intrinsic dimension has to be determined; and third, the
measurement of the magnification from the density of weight vectors is
rather coarse, especially in higher dimensions. 
 \clearpage\noindent
\begin{table}[H]
\small
\caption{\small Comparison of Magnification Control for the Different Control
Approaches for SOM and NG ($d=1$ for SOM).
\label{tab_magnification_control_SOM_NG2}}
\begin{center}
$%
\small
\begin{tabular}{lcc}
\hline
& SOM & NG \\ 
\hline
Local & $\left( m+1\right) \alpha _{SOM}$ & $\left( m+1\right) \alpha _{NG}$
\\ 
learning
& 
(Bauer et al., 
1996)
 & 
(Villmann 2000) 
\\ 
&  &  \\ 
Winner-relaxing & $\frac{3}{\kappa +3}\alpha _{SOM}$ & $\frac{1}{1-\kappa }%
\alpha _{NG}$ \\ 
learning
& 
(Claussen, 2003, 2005) & 
(Claussen \& Villmann, 2005) \\ 
&  &  \\ 
{Concave-convex} & $\frac{3}{\xi +2}\alpha _{SOM}$ & $\frac{d+2}{d+\xi +1}%
\alpha _{NG}$ \\ 
learning & (in section 4.3; & in section 5.3 \\ 
& 
 Zheng \& Greenleaf, 1996)
 &  \\ \hline
\end{tabular}%
$%
\end{center}
\normalsize
\end{table}
\noindent
Only some special cases 
{can be handled adequately. 
}
In particular, max\-imi\-zing mutual information
can be controlled easily by observation of the entropy of winning
probabilities of neurons or consideration of inverted magnification in case
of available auxiliary class information, that is, labeled data 
(Mer{\'e}nyi \& Jain, 2004).
Thus, actual applications have to be done carefully
using some heuristics. Interesting, successful applications of magnification
control (by local learning) in satellite remote sensing image analysis can
be found in 
Mer{\'e}nyi and Jain (2004),
(Villmann, 1999;
Villmann {et al}., 2003).

Summarizing the above approaches of magnification control, we obtain the
good news that the possibilities for magnification control known from SOM
can be successfully transferred to the NG learning in all three cases. The
achieved theoretical magnifications are collected in 
Table \ref{tab_magnification_control_SOM_NG2}.

The interesting point is that the local learning approach, as well as 
{concave-convex
} 
learning, yields structurally similar modification factors for
the new magnification. However, a magnification of $1$ is not reachable
by 
{concave-convex
} 
learning in
case of NG. In case of the winner relaxing approach, we have a remarkable
difference: in contrast to the WRSOM, where the relaxing term has to be
inverted ($\kappa <0$) to increase the magnification exponent, for the
NG, positive values of $\kappa $ are required to increase the
magnification factor.

\subsection*{\normalsize 
Appendix A: 
 Magnification Law of the Generalized
{Concave-Convex
}
Learning for 
the Self-Organizing Map
\label{appendix_general_zheng_som}}
\renewcommand{\theequation}{A.\arabic{equation}}
\setcounter{equation}{0}
In this appendix we prove the magnification law of the generalized
{concave-convex
} 
learning for SOM:
the exponent in equation (\ref%
{som_zheng_learning}) is required to be $\xi \in 
\mathbb{R}
$ and keeping further in mind the definition (\ref{def_vector_potenz}).
Since the convergence proofs of SOM are only valid for the one-dimensional
setting, we switch from $\mathbf{w}$ to $w$ and from $\mathbf{v}$ to $v$.
 \clearpage\noindent

In the continuum approach, we can replace the index of the neuron by its
position or location $r$ (Ritter {et al.}, 1992). Further, the neighborhood
function $h_{\sigma }$ 
{depends only on 
}
the difference of the location $r$ to 
$r_{s\left( v\right) }$ as the location of the winning neuron. Then we have
in the equilibrium for the learning rule 
equation (\ref{som_zheng_learning}), 
\begin{equation}
\int h_{\sigma }\left( r-r_{s\left( v\right) }\right) \left( v-w\left(
r\right) \right) ^{\xi }P\left( v\right) dv=0.
\label{som_generalized_zheng_gleichgewicht}
\end{equation}%
We perform 
the usual approach of expanding 
   the integrand
in a Taylor series in powers
of $\varsigma =s\left( v\right) -r$ and evaluating at $r$ 
(Ritter \& Schulten, 1986;
Hertz, Krogh, \& Palmer, 1991;
Zheng \& Greenleaf, 1996). 
This gives 
\begin{equation}
v=w\left( r+\varsigma \right) ,
\label{som_generalized_zheng_continuum_annahme}
\end{equation}
\vspace*{4mm}
$h_{\sigma }\left( s\left( v\right) -r\right) $ becomes $h_{\sigma }\left(
\varsigma \right) =h_{\sigma }\left( -\varsigma \right)$, and 
\begin{equation}
P\left( v\right) =P\left( w\left( r+\varsigma \right) \right) \approx
P\left( w\right) +\varsigma P^{\prime }\left( w\right) w^{\prime }\left(
r\right) .
\end{equation}%
Further, $dv=dw\left( r+\varsigma \right) =w^{\prime }\left( r+\varsigma
\right) d\varsigma $ can be rewritten as 
\begin{equation}
w^{\prime }\left( r+\varsigma \right) d\varsigma \approx \left( w^{\prime
}+\varsigma w^{\prime \prime }\right) d\varsigma, 
\end{equation}%
and for $v-w\left( r\right) =w\left( r+\varsigma \right) -w\left( r\right) $
we get 
\begin{equation}
w\left( r+\varsigma \right) -w\left( r\right) \approx \varsigma w^{\prime }+%
\frac{1}{2}\varsigma ^{2}w^{\prime \prime }=\varsigma \left( w^{\prime }+%
\frac{1}{2}\varsigma w^{\prime \prime }\right) .
\end{equation}%
Because of $\left( v-w\left( r\right) \right) ^{\xi }$ in 
equation (\ref{som_generalized_zheng_gleichgewicht}), 
we consider $\left( w^{\prime }+\frac{%
1}{2}\varsigma w^{\prime \prime }\right) ^{\xi }$:%
\begin{eqnarray}
\left( w^{\prime }+\frac{1}{2}\varsigma w^{\prime \prime }\right) ^{\xi }
\approx
%
\label{som_generalized_zheng_potenznaeherung_1} 
\left( w^{\prime }\right) ^{\xi }\left( 1+\frac{\varsigma \cdot \xi }{2}%
\frac{w^{\prime \prime }}{w^{\prime }}\right). 
\label{som_generalized_zheng_potenznaeherung_2}
\end{eqnarray}%
Further, because of the definition (\ref{def_vector_potenz}), 
the power $\varsigma ^{\xi }$ has to be interpreted as 
\begin{equation}
\varsigma ^{\xi }=\varsigma \cdot \left\vert \varsigma 
\right\vert^{\xi-1},
\label{def_zeta_power}
\end{equation}%
\vspace*{4mm}
which is an odd function in $\varsigma$.  

Collecting now (\ref{som_generalized_zheng_continuum_annahme})--(\ref%
{def_zeta_power}) we get in (\ref{som_generalized_zheng_gleichgewicht}) 
\vspace*{4mm}
\begin{eqnarray}
0 &=&\int h_{\sigma }\left( \varsigma \right) \cdot \varsigma \cdot
\left\vert \varsigma \right\vert ^{\xi -1}\cdot \left( w^{\prime }\right)
^{\xi }\cdot \left( 1+\frac{1}{2}\xi w^{\prime \prime }\left( w^{\prime
}\right) ^{-1}\varsigma \right)   \label{equ_appendex_h1} \\
&&\times \left( P\left( w\right) +\varsigma P^{\prime }\left( w\right)
w^{\prime }\left( r\right) \right) \left( w^{\prime }+\varsigma w^{\prime
\prime }\right) d\varsigma.   
\notag
\end{eqnarray}%
 \clearpage\noindent
Since $\varsigma \cdot \left\vert \varsigma \right\vert ^{\xi -1}$
is odd, 
the term of lowest order in $\varsigma$ vanishes
according to
the rotational symmetry of $h_{\sigma }\left( \varsigma \right) $. Further,
in our approximation, we ignore terms behind $\varsigma ^{2}$. Hence, the
above equation can be simplified as 
\vspace*{4mm}
\begin{equation}
0=\left( w^{\prime }\right) ^{\xi }\left( P^{\prime }\left( w\right) \left(
w^{\prime }\right) ^{2}+\frac{\xi +2}{2}P\left( w\right) w^{\prime \prime
}\right) \int h_{\sigma }\left( \varsigma \right) \cdot \varsigma ^{2}\cdot
\left\vert \varsigma \right\vert ^{\xi -1}d\varsigma.
\label{equ_appendex_h2}
\end{equation}%
\vspace*{4mm}
From there we get%
\vspace*{4mm}
\begin{equation}
\rho =\left\vert \frac{dr}{dw}\right\vert =P^{\frac{2}{2+\xi }}
\end{equation}%
\vspace*{4mm}
and, hence,%
\vspace*{4mm}
\begin{equation}
\alpha_{{concave/convex}SOM}=\frac{2}{2+\xi},
\end{equation}%
\vspace*{4mm}
which completes the proof.

\subsection*{\normalsize Appendix B:
{Magnification Law of the
} 
Generalized
{Concave-Convex
} 
Learning for Neural Gas \label{appendix_general_zheng_NG}}
\renewcommand{\theequation}{B.\arabic{equation}}
\setcounter{equation}{0}
For the derivation of the magnification for the generalized 
{concave-convex
}
learning in case of magnification-controlled NG, first we have the usual
continuum assumption 
(Ritter {et al.}, 1992). 
The further treatment is in complete
analogy to the derivation of the magnification in the usual NG 
(Martinetz et al., 1993).
Let $\mathbf{r}$ be the difference vector
\vspace*{4mm}
\begin{equation}
\mathbf{r}=\mathbf{v}-\mathbf{w}_{i},
\end{equation}%
The winning rank $k_{i}\left( \mathbf{v},\mathbf{W}\right) $ in the
neighborhood function $h_{\lambda }\left( i,\mathbf{v},\mathbf{W}\right) $
in equation (\ref{h_trn}) 
{depends only on
}
$\mathbf{r}$, therefore, we introduce the
new variable 
\vspace*{4mm}
\begin{equation}
\mathbf{x}\left( \mathbf{r}\right) =\mathbf{\hat{r}\cdot k}_{i}\left( 
\mathbf{r}\right) ^{\frac{1}{d}},
\end{equation}%
which can be assumed as monotonously increasing with $\left\Vert \mathbf{r}%
\right\Vert $. We define the $d\times d$--Jacobian 
\vspace*{4mm}
\begin{equation}
\mathbf{J}\left( \mathbf{x}\right) =\det \left( \frac{{\partial r_{k}}}{{%
\partial x_{l}}}\right) .
\end{equation}
 \clearpage\noindent
Starting from the new learning rule, 
\vspace*{4mm}
\begin{equation}
\bigtriangleup \mathbf{w}_{i}=\epsilon h_{\lambda }\left( i,\mathbf{v},%
\mathbf{W}\right) \left( \mathbf{v}-\mathbf{w}_{i}\right)^{\xi},
\end{equation}%
\vspace*{4mm}
again consider the averaged change,
\vspace*{4mm}
\begin{equation}
\left\langle \bigtriangleup \mathbf{w}_{i}\right\rangle =\int P\left( 
\mathbf{v}\right) h_{\lambda }\left( i,\mathbf{v},\mathbf{W}\right) \left( 
\mathbf{v-w}_{i}\right) ^{\xi }d\mathbf{v}\text{ }.
\label{ng_averaged_change}
\end{equation}%
If $h_{\lambda }\left( i,\mathbf{v},\mathbf{W}\right) $ in 
equation (\ref{h_trn})
rapidly decreases to zero with increasing $\mathbf{r}$, we can replace the
quantities $\mathbf{r}\left( \mathbf{x}\right) $, $\mathbf{J}\left( \mathbf{x%
}\right) $ by the first terms of their respective Taylor expansions around
\vspace*{4mm}
the point $\mathbf{x}=0$, neglecting higher derivatives. We obtain 
\begin{equation}
\mathbf{x}\left( \mathbf{r}\right) =\mathbf{r}\left( \tau _{d}\rho \left( 
\mathbf{w}_{i}\right) \right) ^{\frac{1}{d}}\left( 1+\frac{\mathbf{r}\cdot
\partial _{\mathbf{r}}\rho \left( \mathbf{w}_{i}\right) }{d\cdot \rho \left( 
\mathbf{w}_{i}\right) }+\mathcal{O}\left( \mathbf{r}^{2}\right) \right), 
\end{equation}%
\vspace*{4mm}
which corresponds to 
\vspace*{4mm}
\begin{equation}
\mathbf{r}\left( \mathbf{x}\right) =\frac{\left( 1-\left( \tau _{d}\rho
\left( \mathbf{w}_{i}\right) \right) ^{-\frac{1}{d}}\cdot \frac{\mathbf{x}%
\cdot \partial _{\mathbf{r}}\rho \left( \mathbf{w}_{i}\right) }{d\cdot \rho
\left( \mathbf{w}_{i}\right) }+\mathcal{O}\left( \mathbf{x}^{2}\right)
\right) }{\mathbf{x}\left( \tau _{d}\rho \left( \mathbf{w}_{i}\right)
\right) ^{\frac{1}{d}}}
\end{equation}%
\vspace*{4mm}
with 
\vspace*{4mm}
\begin{equation}
\tau _{d}=\frac{{\pi ^{\frac{d}{2}}}}{{\Gamma \left( \frac{d}{2}+1\right) }}
\end{equation}%
as the volume of a $d$--dimensional unit sphere 
(Martinetz {et al}., 1993). 
We
define $\varphi =\tau _{d}\rho \left( \mathbf{w}_{i}\right) $. Further, we
expand $\mathbf{J}\left( \mathbf{x}\right) $ and obtain%
\vspace*{4mm}
\begin{eqnarray}
\mathbf{J}\left( \mathbf{x}\right)  &=&\left( \mathbf{J}\left( 0\right)
+x_{k}\frac{\partial \mathbf{J}}{\partial x_{k}}+\ldots \right)  \\
&=&\frac{1}{\varphi }\left( 1-\varphi ^{-\frac{1}{d}}\left( 1+\frac{1}{d}%
\right) \cdot \mathbf{x}\cdot \frac{\partial _{\mathbf{r}}\rho }{\rho }%
\right) +\mathcal{O}\left( x^{2}\right) 
\end{eqnarray}%
\vspace*{4mm}
and, hence,%
\vspace*{4mm}
\begin{equation}
\left. \frac{\partial \mathbf{J}}{\partial \mathbf{x}}\right\vert _{\mathbf{x%
}=0}=-\varphi ^{-\left( 1+\frac{1}{d}\right) }\frac{\partial _{\mathbf{r}%
}\rho }{\rho }.
\end{equation}%
\clearpage\noindent
After 
collecting all replacements, (\ref{ng_averaged_change}) becomes 
\vspace*{4mm}
\begin{eqnarray}
\left\langle \bigtriangleup \mathbf{w}_{i}\right\rangle  &=&\epsilon \cdot
\varphi ^{-\frac{\xi }{d}}\int_{\mathcal{D}}d\mathbf{x\;}h_{\lambda }\left( 
\mathbf{x}\right) \cdot \mathbf{x}^{\xi }\cdot   \notag \\
&&\cdot \left( P+\varphi ^{-\frac{1}{d}}\cdot \mathbf{x}\cdot \partial _{%
\mathbf{r}}P+\ldots \right) \cdot  \\
&&\cdot \left( \frac{1}{\varphi }-\left( 1+\frac{1}{d}\right) \varphi
^{-\left( 1+\frac{1}{d}\right) }\cdot \mathbf{x}\cdot \frac{\partial _{%
\mathbf{r}}\rho }{\rho }+\ldots \right) \cdot  \\
&&\cdot \left( 1-\varphi ^{-\frac{1}{d}}\cdot \mathbf{x}\cdot \frac{\partial
_{\mathbf{r}}\rho }{d\cdot \rho }+\ldots \right) ^{\xi},
\end{eqnarray}%
\vspace*{4mm}
with new integration variable $\mathbf{x}$. We use the approximation 
\vspace*{4mm}
\begin{equation}
\left( 1-\varphi ^{-\frac{1}{d}}\cdot \mathbf{x}\cdot \frac{\partial _{%
\mathbf{r}}\rho }{d\cdot \rho }+\ldots \right) ^{\xi }\approx 1-\xi \varphi
^{-\frac{1}{d}}\cdot \mathbf{x}\cdot \frac{\partial _{\mathbf{r}}\rho }{%
d\cdot \rho }+\ldots 
\end{equation}%
\vspace*{4mm}
and get
\vspace*{4mm}
\begin{eqnarray}
\left\langle \bigtriangleup \mathbf{w}_{i}\right\rangle  &=&\epsilon \cdot
\varphi ^{-\frac{\xi }{d}}\int_{\mathcal{D}}d\mathbf{x\;}h_{\lambda }\left( 
\mathbf{x}\right) \cdot \mathbf{x}^{\xi }  \notag \\
&&\cdot \left( P+\varphi ^{-\frac{1}{d}}\cdot \mathbf{x}^{\xi }\cdot
\partial _{\mathbf{r}}P+\ldots \right)  \\
&&\cdot \left( \frac{1}{\varphi }-\left( 1+\frac{1}{d}\right) \varphi
^{-\left( 1+\frac{1}{d}\right) }\cdot \mathbf{x}^{\xi }\cdot \frac{\partial
_{\mathbf{r}}\rho }{\rho }+\ldots \right)  \\
&&\cdot \left( 1-\xi \varphi ^{-\frac{1}{d}}\cdot \mathbf{x}^{\xi }\cdot 
\frac{\partial _{\mathbf{r}}\rho }{d\cdot \rho }+\ldots \right). 
\end{eqnarray}%
\vspace*{4mm}
In the equilibrium $\left\langle \bigtriangleup \mathbf{w}_{i}\right\rangle
\vspace*{4mm}
=0$, we have%
\begin{eqnarray}
0 &=&\int_{\mathcal{D}}d\mathbf{x\;}h_{\lambda }\left( \mathbf{x}\right)
\cdot \mathbf{x}^{\xi }\cdot \left( P+\varphi ^{-\frac{1}{d}}\cdot \mathbf{x}%
^{\xi }\cdot \partial _{\mathbf{r}}P+\ldots \right)   \notag \\
&&\cdot \left( \frac{1}{\varphi }-\left( 1+\frac{1}{d}\right) \varphi
^{-\left( 1+\frac{1}{d}\right) }\cdot \mathbf{x}^{\xi }\cdot \frac{\partial
_{\mathbf{r}}\rho }{\rho }+\ldots \right)  \\
&&\cdot \left( 1-\xi \varphi ^{-\frac{1}{d}}\cdot \mathbf{x}^{\xi }\cdot 
\frac{\partial _{\mathbf{r}}\rho }{d\cdot \rho }+\ldots \right). 
\end{eqnarray}%
Because of the rotational symmetry of $h_{\lambda }$, we can neglect odd
power terms in $\mathbf{x}$. Remaining terms are of even power order. Again,
according to equation (\ref{def_vector_potenz}), we take $\mathbf{x}^{\xi }=\mathbf{%
x\cdot }\left\vert \mathbf{x}\right\vert ^{\xi -1}$, and, hence, $\mathbf{x}%
^{\xi }$ itself acts as an odd term. 
Therefore, only terms containing $%
\mathbf{x}^{\xi +k}$ with odd $k$ contribute. Finally,
considering the
non-vanishing terms and neglecting higher order terms, we find the relation 
\begin{equation}
\frac{\partial _{\mathbf{r}}P}{P\left( \mathbf{w}_{i}\right) }=\frac{%
\partial _{\mathbf{r}}\rho }{\rho }\left( \frac{d}{d+\xi +1}\right), 
\end{equation}%
which is the desired result.
\\\\

\paragraph*{Acknowledgements}
\mbox{}\\
We are grateful to
Barbara Hammer (University Clausthal, Germany) 
for intensive discussions
and comments. 
We also thank an anonymeous reviewer 
for a hint that led us to a more
elegant proof in appendix A. 
\\\\

\section*{References}
\begin{description}
\small
\footnotesize
\item[]
Ahalt, S.\ C., Krishnamurty, A.\ K.\, Chen, P., \& Melton, D.\ E.\
(1990).
\newblock
Competitive learning algorithms for vector quantization.
\newblock {\sl Neural Networks}, 3,
277--290.
\\[-4.5ex]

\item[]
Amari, S.-I.\ (1980).
\newblock Topographic organization of nerve fields.
\newblock {\sl Bulletin of Mathematical Biology}, 42, 339--364.
\\[-4.5ex]

\item[]
Bauer, H.\ U., Der, R., \& Herrmann M.\ (1996).
\newblock Controlling the magnification factor of self--organizing feature
  maps.
\newblock {\sl Neural Computation}, 8,
757--771.
\\[-4.5ex]

\item[]
Bauer, H.\ U., Der, R., \& Villmann, Th.\ (1999).
\newblock Neural maps and topographic vector quantization.
\newblock {\sl Neural Networks}, 12,
659--676.
\\[-4.5ex]

\item[]
Bauer, H.\ U., \& Pawelzik, K.\ R.\ (1992).
\newblock Quantifying the neighborhood preservation of {S}elf-{O}rganizing
  {F}eature {M}aps.
\newblock {\sl IEEE Trans. on Neural Networks}, 3,
570--579.
\\[-4.5ex]

\item[]
Bishop, C.\ M., Svens{\'e}n, M., \& Williams, C.\ K.\ I.\ (1998).
\newblock {GTM}: The generative topographic mapping.
\newblock {\sl Neural Computation}, 10, 215--234.
\\[-4.5ex]

\item[]   
Brause, R.\ (1992).
\newblock Optimal information distribution and performance in   neighbourhood-conserving maps for robot control.
\newblock {\sl Int. J. Computers and Artificial Intelligence}, 11,
173--199.
\\[-4.5ex]

\item[]
Brause, R.\ W.\ (1994).
\newblock An approximation network with maximal transinformation.
\newblock In Maria Marinaro and Pietro~G. Morasso, editors, {\sl Proc.
  ICANN'94, International Conference on Artificial Neural Networks},
 vol.\ I,
  pp.\ 701--704, Springer, London.
\\[-4.5ex]

\item[] 
Bruske, J., \& Sommer, G.\ (1998).
\newblock Intrinsic dimensionality estimation with optimally topology
  preserving maps.
\newblock {\sl IEEE Transactions on Pattern Analysis and Machine Intelligence},
  20,
572--575.
\\[-4.5ex]

\item[] 
Camastra, F.\ \& Vinciarelli, A.\ (2001).
\newblock Intrinsic dimension estimation of data: an approach based
  {G}rassberger-{P}rocaccia's algorithm.
\newblock {\sl Neural Processing Letters}, 14, 27--34.
\\[-4.5ex]

\item[] 
Claussen J.~C.\ (2003).
\newblock Winner-relaxing and winner-enhancing Kohonen maps: 
Maximal mutual
  information from enhancing the winner.
\newblock {\sl Complexity}, 
8(4), 15--22.
\\[-4.5ex]

\item[]
Claussen J.~C.\ (2005).
\newblock Winner-Relaxing Self-Organizing Maps.
\newblock {\sl Neural Computation}, 17, 997--1009.
\\[-4.5ex]

\clearpage

\item[]
Claussen J.~C.\ \& H.~G. Schuster, H.~G.\ (2002).
\newblock Asymptotic level density of the Elastic Net 
self-organizing feature
  map.
\newblock In J.R. Dorronsoro, editor, {\sl Proc. International Conf. on
  Artificial Neural Networks (ICANN)}, Lecture Notes in Computer Science 2415,
  pages 939--944. Springer Verlag.
\\[-4.5ex]

\item[]
Claussen J.~C.\ \&  Villmann, Th.\ (2003a).
\newblock Magnification control in winner relaxing neural gas.
\newblock In M.~Verleysen, editor, {\sl Proc. Of European Symposium on
  Artificial Neural Networks{(ESANN'2003)}}, pages 93--98, 
 d-side,
Brussels, Belgium.
\\[-4.5ex]

\item[]
Claussen J.~C.\ \&  Villmann, Th.\ (2003b).
\newblock Magnification control in neural gas by winner relaxing learning:
  Independence of a diagonal term.
\newblock In Kaynak, O., Alpaydin, E., Oja, E., and
Xu, L. editors, {\sl Proc. International Conference on
  Artificial Neural Networks {(ICANN/ICONIP 2003)}}, pages 58--61, 
Bogazici University, 
Istanbul.
\\[-4.5ex]

\item[] 
Claussen J.~C.\  \& Villmann, Th.\ (2005).
\newblock Magnification control in winner-relaxing neural gas.
\newblock {\sl Neurocomputing}, 63,
125--137.
\\[-4.5ex]

\item[] 
Cottrell, M., Fort, J.\ C., \& Pages, G.\ (1998).
\newblock Theoretical aspects of the {SOM} algorithm.
\newblock {\sl Neurocomputing}, 21,
119--138.
\\[-4.5ex]

\item[] 
de Bodt, E., Cottrell, M., Letremy, P., \& Verleysen, M.\ (2004). 
\newblock On the use of self-orgainzing maps to accelerate vector quantization.
\newblock {\sl Neurocomputing}, 17, 187--203.
\\[-4.5ex]

\item[] 
Der, R., \& Herrmann, M.\ (1992).
\newblock Attention based partitioning.
\newblock In M.~Van der Meer, editor, {\sl Bericht des Status--Seminar des BMFT
  Neuroinformatik}, pages 441--446. DLR (Berlin).
\\[-4.5ex]

\item[] 
Dersch, D., \& Tavan., P.\ (1995).
\newblock Asymptotic level density in topological feature maps.
\newblock {\sl IEEE Trans. on Neural Networks}, 6,
230--236.
\\[-4.5ex]

\item[] 
DeSieno, D.\ (1988).
\newblock Adding a conscience to competitive learning.
\newblock In {\sl Proc. ICNN'88, International Conference on Neural Networks},
  pages 117--124, Piscataway, NJ, 1988. IEEE Service Center.
\\[-4.5ex]

\item[] 
Duda, R.\ O., \& Hart, P.\ E.\ (1973).
\newblock {\sl Pattern Classification and Scene Analysis}.
\newblock Wiley, New York, 1973.
\\[-4.5ex]

\item[] 
Durbin, R.\,  \& D.~Willshaw, D.\ (1987)
\newblock An analogue approach to the travelling salesman problem using an
  elastic net method.
\newblock {\sl Nature}, 326, 689--691.
\\[-4.5ex]

\item[] 
Eckmann, J.\ P., \& Ruelle, D.\ (1992).
\newblock Fundamental limitations for estimating dimensions and {L}yapunov
  exponents in dynamical systems.
\newblock {\sl Physica D}, 56, 185--187.
\\[-4.5ex]

\item[] 
Erwin, E., Obermayer, K., \& Schulten, K.\ (1992).
\newblock Self-organizing maps: Ordering, convergence properties and energy
  functions.
\newblock {\sl Biol. Cyb.}, 67,
47--55.
\\[-4.5ex]

\item[] 
Fritzke, B.\ (1993).
\newblock Vector quantization with a growing and splitting elastic net.
\newblock In Stan Gielen and Bert Kappen, editors, {\sl Proc. ICANN'93,
  International Conference on Artificial Neural Networks}, pages 580--585,
Springer, London, UK.
\\[-4.5ex]

\item[] 
Galanopoulos, A.\ S., \& Ahalt, S.\ C.\ (1996).
\newblock Codeword distribution for frequency sensitive competitive learning
  with one dimensional input data.
\newblock {\sl IEEE Transactions on Neural Networks}, 
7,  752--756.
\\[-4.5ex]

\item[] 
Grassberger, P.\  and Procaccia, I.\ (1983).
\newblock Measuring the strangeness of strange attractors.
\newblock {\sl Physica D}, 9, 189--208.
\\[-4.5ex]

\item[] 
Hammer, B., \& Villmann, Th.\ (2003).
\newblock Mathematical aspects of neural networks.
\newblock In M.~Verleysen, editor, {\sl Proc. Of European Symposium on
  Artificial Neural Networks {(ESANN'2003)}}, pages 59--72, 
d-side, Brussels, Belgium.
\\[-4.5ex]

\item[] 
Haykin, S.\ (1994).
\newblock {\sl Neural Networks - A Comprehensive Foundation}.
\newblock IEEE Press, New York.
\\[-4.5ex]

\item[] 
Herrmann, M., Bauer, H.-U., \& Der, R.\ (1994).
\newblock The 'perceptual magnet' effect: A model based on self-organizing
  feature maps.
\newblock In L.~S. Smith and P.~J.~B. Hancock, editors, {\sl Neural Computation
  and Psychology}, pages 107--116, Stirling, Springer-Verlag.
\\[-4.5ex]

\clearpage

\item[] 
Herrmann, M., \& Villmann, Th.\ (1997).
\newblock Vector quantization by optimal neural gas.
\newblock In
W.\ Gerstner, A. Germond, M.\ Hasler, \& J.-D.\ Nicoud (eds.),
 {\sl Artificial Neural Networks -- Proceedings of
  International Conference on Artificial Neural Networks (ICANN'97) Lausanne},
  pages 625--630. Lecture Notes in Computer Science 1327, Springer Verlag,
  Berlin Heidelberg.
\\[-4.5ex]

\item[]
Hertz, J.\ A., Krogh, A., \& Palmer, R.\ G.\ (1991).
\newblock {\sl Introduction to the Theory of Neural Computation}, volume~1 of
  {\sl {S}anta {F}e {I}nstitute Studies in the Sciences of Complexity: Lecture
  Notes}.
\newblock Addison-Wesley, Redwood City, CA.
\\[-4.5ex]

\item[] 
Heskes, T.\ (1999).
\newblock Energy functions for \mbox{self-organizing} maps.
\newblock In E.~Oja and S.~Kaski, editors, {\sl Kohonen Maps}, pages 303--316.
  Elsevier, Amsterdam.
\\[-4.5ex]

\item[]
Jain, A.\ \& Mer{\'e}nyi, E.\ (2004).
\newblock Forbidden magnification? {I}.
\newblock In M.~Verleysen, editor, {\sl European Symposium on Artificial Neural
  Networks 2004}, pages 51--56. d-side publications.
\\[-4.5ex]

\item[] 
Kohonen, T.\ (1991).
\newblock Self-{O}rganizing {M}aps: {O}ptimization approaches.
\newblock In T.~Kohonen, K.~M{\"{a}}kisara, O.~Simula, and J.~Kangas, editors,
  {\sl Artificial Neural Networks}, volume~II, pages 981--990, Amsterdam,
  Netherlands, 1991. North-Holland.
\\[-4.5ex]

\item[] 
Kohonen, T.\ (1995).
\newblock {\sl {Self-Organizing Maps}}, volume~30 of {\sl Springer Series in
  Information Sciences}.
\newblock Springer, Berlin, Heidelberg.
\newblock (Second Extended Edition 1997).
\\[-4.5ex]

\item[] 
Kohonen, T.\ (1999).
\newblock Comparison of SOM point densities based on different criteria.
\newblock {\sl Neural Computation}, 11, 2081--95.
\\[-4.5ex]

\item[] 
Kuhl, P.\ K.\ (1991).
\newblock Human adults and human infants show a 'perceptual magnet' effect for
  the prototypes of speech categories, monkeys do not.
\newblock {\sl Perception and Psychophysics},
50, 93--107.
\\[-4.5ex]

\item[] 
Kuhl, P.\ K., Williams, K.\ A., Lacerda, F.,  Stevens, K.\ N.\ \& Lindblom, B.\
(1992).
\newblock Linguistic experience alters phonetic perception in infants by 6
  months of age.
\newblock {\sl Science}, 255, 606--608.
\\[-4.5ex]

\item[] 
Liebert, W.\ (1991).
\newblock {\sl Chaos und Herzdynamik}.
\newblock Verlag Harri Deutsch, Frankfurt/M., Germany.
\\[-4.5ex]

\item[] 
Linde, Y., Buzo, A., \& Gray, R.\ M.\ (1980).
\newblock An algorithm for vector quantizer design.
\newblock {\sl IEEE Transactions on Communications}, 28, 84--95.
\\[-4.5ex]

\item[] 
Linsker, R.\ (1989).
\newblock How to generate maps by maximizing the mutual information between
  input and output signals.
\newblock {\sl Neural Computation}, 1, 402--411.
\\[-4.5ex]

\item[] 
Luttrell, S.\ P.\ (1991).
\newblock Code vector density in topographic mappings: scalar case.
\newblock {\sl IEEE Trans. on Neural Networks}, 2,
427--436.
\\[-4.5ex]

\item[] 
Martinetz, Th.\ M., Berkovich, S.\ G., \& Schulten, K.\ J.\ (1993).
\newblock ``{N}eural-gas'' network for vector quantization and its application to
  time-series prediction.
\newblock {\sl {IEEE} Trans. on Neural Networks}, 4,
558--569.
\\[-4.5ex]

\item[] 
Martinetz, Th., \& Schulten, K.\ (1994).
\newblock Topology representing networks.
\newblock {\sl Neural Networks}, 7,
507--522.
\\[-4.5ex]

\item[] 
Mer{\'e}nyi, E.,  \& Jain, A.\ (2004).
\newblock Forbidden magnification? {II}.
\newblock In M.~Verleysen, editor, {\sl European Symposium on Artificial Neural
  Networks 2004}, pages 57--62. d-side publications.
\\[-4.5ex]

\item[] 
Ripley, B.\ D.\ (1996).
\newblock {\sl Pattern Recognition and Neural Networks}.
\newblock Cambridge University Press.
\\[-4.5ex]

\item[] 
Ritter, H.\ (1989).
\newblock Asymptotic level density for a class of vector quantization
  processes.
\newblock Report~A9, Helsinki University of Technology, Laboratory of Computer
  and Information Science, Espoo, Finland.
\\[-4.5ex]

\item[] 
Ritter, H.\ (1991).
\newblock Asymptotic level density for a class of vector quantization
  processes.
\newblock {\sl IEEE Trans. on Neural Networks},
 2,
173--175.
\\[-4.5ex]

\item[] 
Ritter, H., \& Schulten, K.\ (1986).
\newblock On the stationary state of {{K}ohonen's} self-organizing sensory
  mapping.
\newblock {\sl Biol. Cyb.}, 54, 99--106.
\\[-4.5ex]

\clearpage

\item[] 
Ritter, H., Martinetz, Th., \& Schulten, K.\
(1992).
\newblock {\sl Neural {C}omputation and {S}elf-{O}rganizing {M}aps: {A}n
  {I}ntroduction}.
\newblock Addison-Wesley, Reading, MA.
\\[-4.5ex]

\item[] 
Takens, F.\ (1985).
\newblock On the numerical determination of the dimension of an attractor.
\newblock In B.~Braaksma, H.~Broer, and F.~Takens, editors, {\sl Dynamical
  Systems and Bifurcations}, pages 99--106, 
Springer, Berlin.
\\[-4.5ex]

\item[] 
Theiler, J.\ (1990).
\newblock Satistical precision of dimension estimators.
\newblock {\sl Physical Review A}, 41, 3038--3051.
\\[-4.5ex]

\item[] 
van~Hulle, M.\ M.\ (2000).
\newblock {\sl Faithful Representations and Topographic Maps From Distortion-
  to Information-based Self-organization}.
\newblock J. Wiley \& Sons, Inc.
\\[-4.5ex]

\item[] 
 Villmann, Th.\ (1999).
\newblock Benefits and limits of the self-organizing map and its variants in
  the area of satellite remote sensoring processing.
\newblock In {\sl Proc. Of European Symposium on Artificial Neural Networks
  (ESANN'99)}, pages 111--116, 
D facto publications,
Brussels, Belgium. 
\\[-4.5ex]

\item[] 
Villmann, Th.\ (2000).
\newblock Controlling strategies for the magnification factor in the neural gas
  network.
\newblock {\sl Neural Network World}, 
10,
739--750.
\\[-4.5ex]

\item[] 
Villmann, Th.\ (2002).
\newblock Neural maps for faithful data modelling in medicine -- state of the
  art and exemplary applications.
\newblock {\sl Neurocomputing}, 48,
229--250.
\\[-4.5ex]

\item[] 
Villmann, Th., \& Heinze, A.\ (2000).
\newblock Application of magnification control for the neural gas network in a
  sensorimotor architecture for robot navigation.
\newblock In Horst-Michael Gro{\ss}, Klaus Debes, and Hans-Joachim B{\"o}hme,
  editors, {\sl Proceedings of Selbstorganisation Von Adaptivem Verfahren
  {(SOAVE'2000)} Ilmenau}, pages 125--134, VDI-Verlag D\"usseldorf, 2000.
  Fortschrittsberichte des VDI.
\\[-4.5ex]

\item[] 
Villmann, Th., Hermann, W., \& Geyer, M.\ (2000).
\newblock Variants of self-organizing maps for data mining and data
  visualization in medicine.
\newblock {\sl Neural Network World}, 
10,
751--762.
\\[-4.5ex]

\item[] 
Villmann, Th., Mer{\'e}nyi, E., \& Hammer, B.\ (2003).
\newblock Neural maps in remote sensing image analysis.
\newblock {\sl Neural Networks}, 16,
389--403.
\\[-4.5ex]

\item[] 
Willshaw, D.\ J.,  \&  von der Malsburg, C.\ (1976).
\newblock How patterned neural connections can be set up by self--organization.
\newblock {\sl {Proceedings of the Royal Society of London, Series B}},
  194,
431--445.
\\[-4.5ex]

\item[] 
Zador, P.\ L.\ (1982).
\newblock Asymptotic quantization error of continuous signals and the
  quantization dimension.
\newblock {\sl IEEE Transaction on Information Theory}, 
28,149--159.\\[-4.5ex]


\item[] 
Zheng, Y., \& Greenleaf, J.~F.\ (1996).
\newblock The effect of concave and convex weight adjustments on
  \mbox{self-organizing} maps.
\newblock {\sl IEEE Transactions on Neural Networks}, 7, 87--96.
\\[-5.5ex]
\end{description}
\underline{~~~~~~~~~~~~~~~~~~~~~~~~~~~~~~~~}\\[-1mm]
{\small\footnotesize 
Received June 15, 2004; accepted June 15, 2005.}

\end{document}